# Phase Coexistence Near a Morphotropic Phase Boundary in Sm-doped BiFeO$_3$ Films

S.B. Emery[1], C.-J. Cheng[2], D. Kan[3], F.J. Rueckert[1], S.P. Alpay[1,4], V. Nagarajan[2], I. Takeuchi[3], and B.O. Wells[1]

[1] *Department of Physics, University of Connecticut, Storrs, Connecticut 06269, USA*

[2] *School of Materials Science and Engineering, University of New South Wales, New South Wales 2052, Australia*

[3] *Department of Materials Science and Engineering, University of Maryland, College Park, Maryland 20742, USA*

[4] *Materials Science and Engineering Program and Institute of Materials Science, University of Connecticut 06269, USA*

## Abstract

We have investigated heteroepitaxial films of Sm-doped BiFeO$_3$ with a Sm-concentration near a morphotropic phase boundary. Our high-resolution synchrotron X-ray diffraction, carried out in a temperature range of 25°C to 700°C, reveals substantial phase coexistence as one changes temperature to crossover from a low-temperature PbZrO$_3$-like phase to a high-temperature orthorhombic phase. We also examine changes due to strain for films exhibiting anisotropic misfit between film and substrate. Additionally, thicker films exhibit a substantial volume collapse associated with the structural transition that is suppressed in thinner films.

Bismuth ferrite (BiFeO$_3$, BFO) is a material being intensively investigated as it possesses a simple ABO$_3$ rhombohedral distorted-perovskite structure (R3c) with an established tilt system[1] together with two robust ferroic properties at room temperature, i.e., ferroelectricity with a Curie temperature ($T_C$) of ~ 830°C and an anti-ferromagnetic ordering with a Néel temperature ($T_N$) of ~ 380°C.[2] It also has a notable piezoelectric response making it a possible candidate for replacing lead based piezoelectric materials. However, some drawbacks for BFO include large coercive fields, high-leakage currents, and extremely stringent deposition parameters that have hampered the development of BFO epitaxial thin films applications. These issues can be reduced via the chemical substitution of rare earth elements onto *A*-sites in the BFO lattice.[3-6] Doping significantly improves the properties of these materials and may open more opportunities for designing nanoelectronic devices with improved functional properties.

Recent comprehensive investigations of Sm-doped BFO (BSFO) thin films over various compositions discovered a critical composition of 14% Sm-concentration at room temperature.[6] At this critical doping, the film has features of a morphotropic phase boundary (MPB) solid solution, consisting of a cell-doubled orthorhombic Pnma phase beyond the MPB and a PbZrO$_3$ (PZO)-like structural phase in local regions below the MPB.[6-8] Electron and laboratory X-ray diffraction studies in Ref. [7] show a superlattice peak at ½-order along *b* in pseudo-cubic notation that appears to be uniquely associated with the orthorhombic phase. In this same work, high resolution transmission electron microscopy (HR-TEM) imaging has revealed nanoscale twin domains with a size of 10-20nm. In this Letter, we report on high resolution synchrotron X-ray diffraction that shows substantial phase coexistence as one moves through the temperature regime to cross over from the PbZrO$_3$-like phase to the orthorhombic phase. We also observe anti-phase domain boundaries in the orthorhombic phase that may correspond to nanodomains identified in HR-TEM.[7] Finally, we note that relatively thicker films (200 nm) exhibit a substantial volume collapse associated with the structural transition that is suppressed in thin films (20 nm).

Three films of 11 % BSFO with thicknesses of 20 nm, 100 nm, and 200 nm were grown on cubic (001) SrTiO$_3$ (STO) substrates by pulsed laser deposition using a KrF laser. The growth conditions and parameters are detailed elsewhere.[6] Wavelength dispersive X-ray spectroscopy (WDS) maps over a 100 micron by 100 micron area showed homogenous composition within ± 1%, and extensive TEM studies using HR-TEM and High Angle Annular Dark Field STEM (HAADF-STEM) have not revealed macroscale composition inhomogeneity.[7,9] This composition allows us to straddle the MPB in temperature space and examine in detail the full evolution of the structural features because of a finite slope of the MPB in BSFO (the black line in Fig. 1(a)) as one can see in the temperature-composition phase diagram[10] depicted in Fig. 1(a). We performed X-ray scattering studies at the X20A bending magnet beamline at the National Synchrotron Light Source, Brookhaven National Laboratory. X-ray incident energy was set to 8.04724 keV, with either vertical slits (2 mm) or a Ge(111) analyzer crystal used define the resolution. Absolute $2\theta$ angles were determined using the STO substrate peaks as a reference, and then corrected for temperature changes to sample height using the STO thermal expansion coefficient.[11] In this study, we list all diffraction peaks as (h k l)$_C$//(h' k' l')$_O$, with the former being in pseudo-cubic notation with the cubic $c$-axis normal to the substrate and the latter as orthorhombic (*Pnma*) notation with the orthorhombic $b$-axis in the plane, and orthorhombic $a$ and $c$ at 45° to the substrate ((101)$_O$ orientation).[12] Complete sets of data have been collected for all three films; however, we focus on the thinnest and thickest films to emphasize the role of epitaxial strain. The data for the 100 nm film is consistent with that of the 200 nm film as expected, since both are significantly thicker than the critical thickness for misfit dislocation formation.

Fig. 1(b,c) presents scans in the cubic $h$ direction of the orthorhombic cell-doubled $(0½2)_C//(212)_O$ superstructure peaks as a function of temperature from 25°C to 700°C. Diffraction profiles are offset vertically for clarity, but are otherwise normalized with respect to each other. In general, we find three peaks along this direction, a central peak, and two satellite peaks that have split symmetrically from the central peak. Scans in the cubic $k$ and $l$ directions reveal only a single peak connected to cell-doubling along the b-axis. The split peak detected in the $h$ direction, however, is indicative of a long range distortion of the octahedral tilting pattern that is orthogonal to the long axis. Below 250°C the central peak is more prominent than the satellite peaks in both films. Above this temperature the satellite peaks grow, and become more prominent than the central peak. In both films an approximate crossover temperature of 250°C is found. The trends allow us to label the central peak as being associated with the low temperature phase, whereas the satellite peaks are associated with a high temperature phase. Though the trend is that the central peak dominates at low temperature and the satellites dominate at high temperature, in fact all peaks remain through out the entire temperature range. Previous work did not reveal the weak peak at this position in the low temperature phase and were not of sufficient resolution to distinguish the splitting.[7] The differences in film thickness manifest themselves in two ways: in the magnitude of the splitting of the satellite peaks and in the ratio of intensities of the peaks associated with each phase. Splitting about the h=0 value, coupled to the lack of such splitting in the main Bragg peaks, is not consistent with orthorhombic twin domains but more likely associated with an anti-phase domain boundary of $FeO_6$ octahedra rotations. We measure a periodicity of ~ 33 and ~ 70 pseudo-cubic unit cells in the 200 nm and 20 nm films,

respectively, corresponding to domain sizes of ~ 13 nm and ~ 28 nm. Additionally, temperature cycling to up to 700°C has irreversibly changed the thinner sample. The reason for this is most likely oxygen and/or bismuth desorption, resulting in a change of the stoichiometry in the film.

Fig. 2 shows the primary structural peaks for the two films from Fig. 1 (b,c). The peaks are $(002)_C//(202)_O$, $(012)_C//(222)_O$, and $(112)_C//(321)_O$; here the $(112)_C//(321)_O$ peak is twinned and the $(\bar{1}12)_C//(123)_O$ is also present. The structural peaks also show the phase crossover at 250° C, where the dominant peaks in the spectra change at this temperature. This crossover occurs in all the peaks, but it is difficult to see in the $(112)_C//(321)_O$ peak because of little change in the Q-values and the presence of twinning in both phases. In Fig. 2(a-c) we note that above the crossover temperature the peaks appear at higher $2\theta$ angles and therefore at larger Q in plane. Twinning of only the $(112)_C/(321)_O$ peak is consistent with an orthorhombic phase with the b-axis in-plane.

Lattice parameters have been derived from structural peaks in Fig. 2, and are reported for the orthorhombic cell. Plotted in Fig. 3(a,b) are the derived parameters displayed in a way to indicate the crossover of the low temperature phase (solid symbols) to the high temperature phase (open symbols). The high temperature phase in the thick film, Fig. 3(a), exhibits smaller lattice constants than the low temperature phase, which is consistent with the presence of peaks at larger Q values. In the thinner film there is no such drop in the lattice parameters. Consistent with these findings, there is a volume loss of ~ 2% in the thick film but only ~ 0.2% in the thin film. Only for the *b* parameter can we accurately track lattice constants for both the high temperature and low temperature regimes. While we can follow two peaks in $(002)_C$ and $(012)_C$ regions, the broadness of the $(112)_C//(321)_O$ peak with two phases, both twinned, makes it difficult to track the lattice constants for the minority phases.

Emery *et al.*, APL 5

The measured lattice constants of the both the thin and thicker film indicate that the in-plane strain state is anisotropic. At 700°C, the lattice parameters of the film in the pseudo-cubic setting demonstrate that in both the 20 nm and 200 nm films the *b* parameter is ~ 0.394 nm is a near match to STO with a constant of ~ 0.393 nm. Thus all of the films grow near coherently with respect to the *b* axis. However, relaxation occurs along the other in-plane direction, which is a combination of the orthorhombic *a* and *c* projections onto the film-substrate interface, or the pseudo-cubic a parameter. We find these lattice constants to be ~ 0.395 nm and ~ 0.401 nm for 20 nm and 200 nm films respectively. Since there is a limited amount of bulk and/or single-crystal data on BSFO, a quantitative description of the formation of misfit (interfacial) dislocations is difficult. Nonetheless, it is clear from the high temperature lattice constants that there is a thickness dependent variation in the strain state between 20 nm and 200 nm thick films. Thus the thinner film is highly strained along the pseudo-cubic a direction while the thicker film is likely to be fully relaxed at the growth temperature.

Work on bulk ceramics of BSFO by Karimi *et al.* [12] showed that the temperature-doping phase diagram for the material is such that as one increases the temperature for 13% Sm the sample goes from a phase with a structure similar to $PbZrO_3$ (*Pbam*) to an orthorhombic phase common for orthoferrites (*Pnma*) at ~ 250°C. Our measurements on films, both of the lattice constants and the superlattice peak, are consistent with this basic structure determination. The satellite reflections about the $(0½2)_C//(212)_O$ superstructure peak are narrow, and gain in intensity above 250°C suggesting that the sample is favoring a structural phase that is more predominantly the *Pnma* structure. However, in the low-temperature regime the peaks associated with the high temperature phase are still detectable and in the high temperature regime the peaks associated with the low temperature structure are detectable. For the thicker film, the central peak (low-temperature phase) is more prominent while for the thinner film the satellite peaks (high-temperature phase) dominate.

The presence of both satellite peaks and the central superlattice peak at all temperatures, as well as the observation of the multiple main Bragg peaks, lead us to believe that this system exhibits phase coexistence over a 25°C to 700°C temperature range. The findings of a phase crossover like we have observed, are not typical of a simple polymorphic phase transition. This evidence is consistent with a nanodomain driven multiphase picture at a MPB as observed in lead zirconate titanate (PZT).[13-14] Phase coexistence via the formation of nanodomains is not surprising as TEM results have found local evidence for it,[7] and it appears to be a common feature in PZT at the proximity of a MPB.[15] Moreover, recent calculations based on fundamental thermodynamic principles provide a strong theoretical support for this phenomenon.[15-16] Varying the thickness of the film allows us to tune to different strain states to examine more closely the effects the nanodomains have on the structure. For thick films in the fully relaxed state, the lattice parameters favor more phase coexistence above the crossover temperature with smaller domain sizes. In highly strained films, the high-temperature phase is strongly favored and the sizes of domains are increased. We note that in both cases the amount of strain has at best a weak effect on the crossover temperature.

In summary, using high resolution X-ray diffraction, we have observed phase coexistence over a substantial temperature range (25-700°C) for both thick and thin films of BSFO near a MPB. This seems to support models that show that a MPB can be thought of as consisting of coexisting nanodomains. In addition, we have shown that by varying the film thickness we can alter the strain state, the superstructure, and the ratio of phases present in the films.

Research at the University of Connecticut was supported by NSF grant DMR-0907197. Research at UNSW was supported by an ARC Discovery Project, NEDO and a DEST ISL grant. Work at Maryland was supported by access to the Shared Experimental Facilities of the UMD-NSF-MRSEC (DMR 0520471), NSF DMR 0603644, and ARO W911NF-07-1-0410.

**Figure Captions**

**Figure 1: (a)** Phase diagram for BSFO, where the solid black line is the MPB between the rhombohedral (light blue) and the orthorhombic (red) phases. The PZO-like structure is locally observed in the composition region in dark blue. The dashed line marks the 11% Sm composition. **(b-c)** Temperature dependent scans along h in reciprocal space over the pseudo-cubic superlattice position $(0½2)_C//(212)_O$. Lines were drawn to guide the eye. **(b)** is the 200 nm film where its satellites can be seen developing from the central peak during heating and **(c)** is the 20 nm film showing a distinct loss of the central peak until a chemical change occurs above 650°C.

**Figure 2:** Temperature evolution of several structural peaks on the 200 nm and 20 nm samples. Lines were added as guides to the eye. **(a & d)** are the $(002)_C//(202)_O$ peak for 200nm and 20nm films, respectively, **(b & e)** are the $(012)_C//(222)_O$ peak of the 200nm and 20nm films, respectively, and, **(c & f)** are the $(112)_C//(321)_O$ of the 200nm and 20nm films, respectively.

**Figure 3:** Lattice parameters of BSFO thin films on cubic (001) STO. The lattice parameters are calculated in terms of the orthorhombic unit cell from the structural peaks displayed in Fig. 2. **(a)** and **(b)** are the orthorhombic lattice parameters of the 200 nm and 20 nm films respectively. Here the circles are the cell-doubled *b* parameter, and the squares and triangles are the *c* and *a* lattice parameters, respectively. The open and closed symbols represent the high- and low-temperature phases, respectively.

**Figure 1**

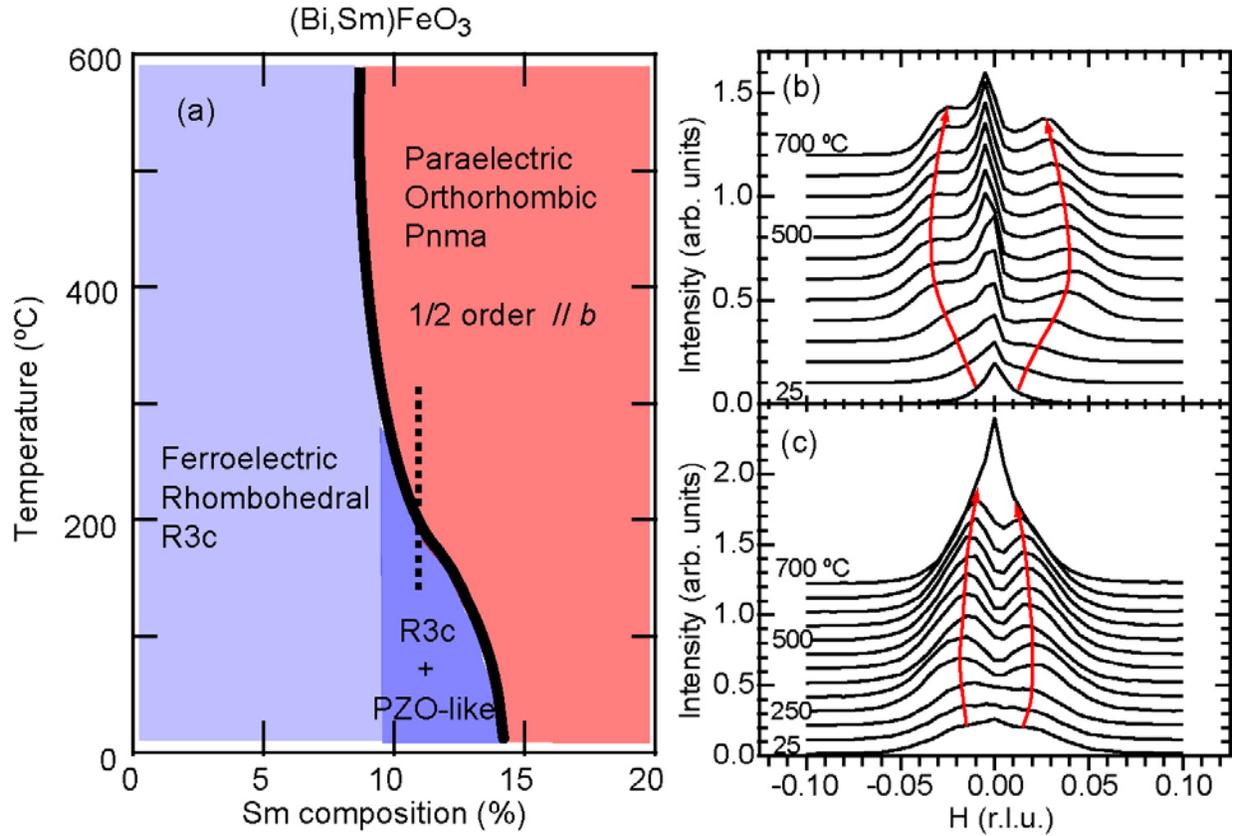

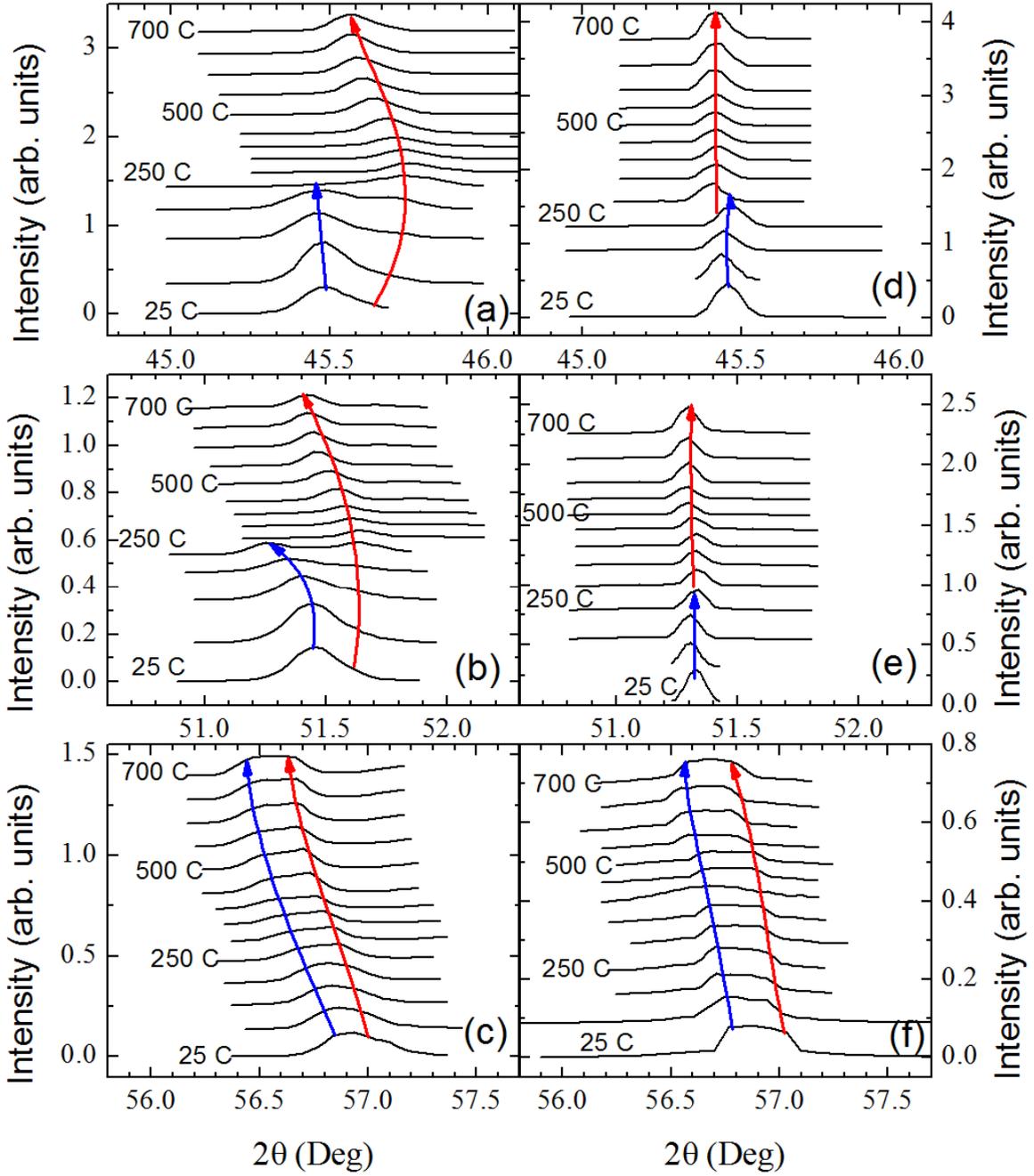

**Figure 3**

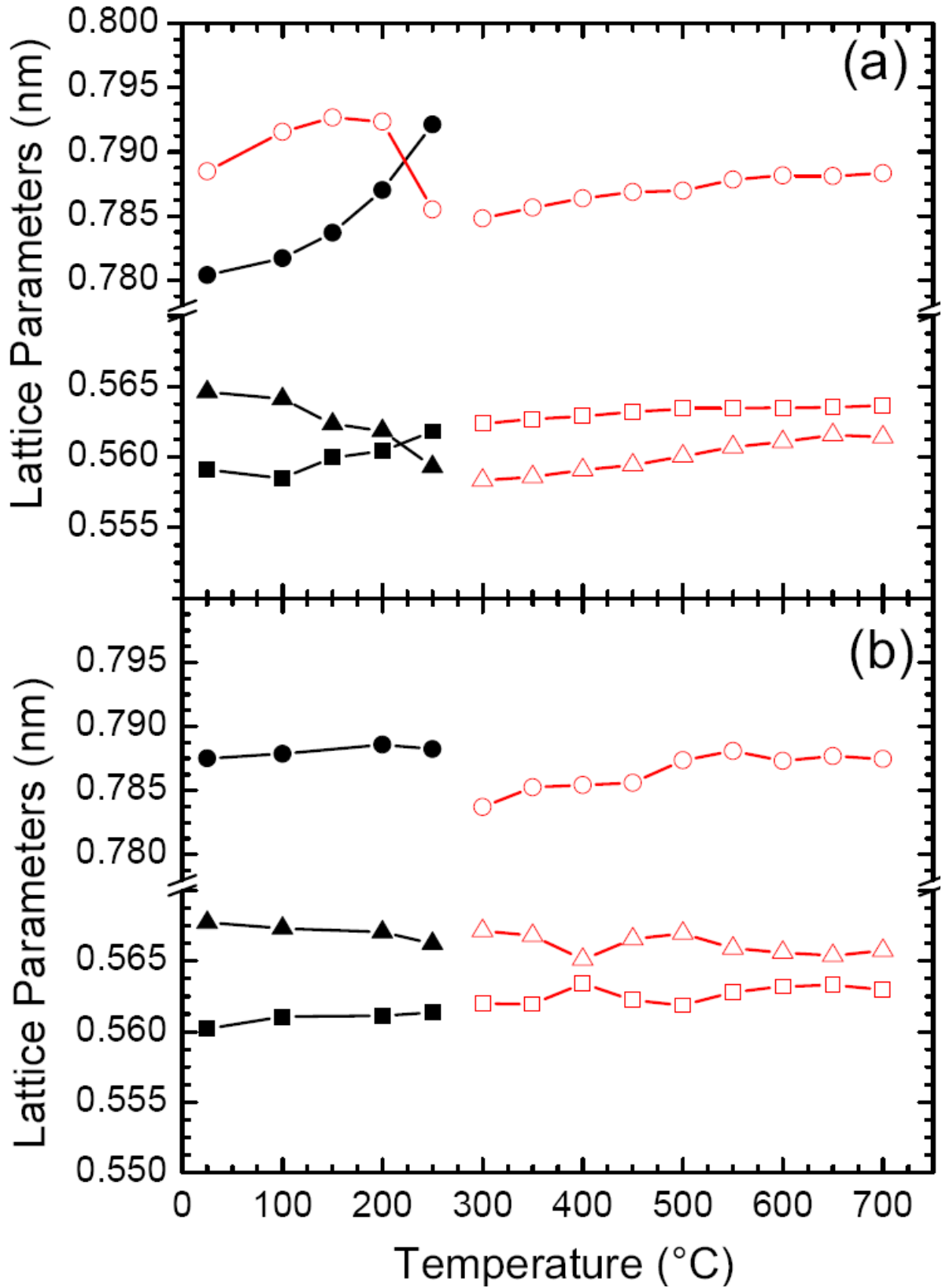